\documentclass[aps, prb, reprint, amsmath, amssymb, longbibliography, superscriptaddress]{revtex4-2}

\usepackage[T1]{fontenc}
\usepackage{palatino}
\usepackage{mathpazo}
\usepackage{soul}
\soulregister\cite7
\soulregister\ref7
\soulregister\pageref7
\soulregister\ce7
\usepackage{graphicx}
\usepackage[svgnames]{xcolor}
\usepackage{amsmath, bm}
\usepackage[detect-none]{siunitx}
\usepackage{fancyref}
\usepackage[colorlinks,
            linkcolor=DarkGreen,
            urlcolor=Blue,
            citecolor=DarkGreen]{hyperref}
\usepackage[utf8]{inputenc}
\usepackage{textcomp}
\usepackage[version=3]{mhchem}
\usepackage{miller}

\newcommand{\orcidicon}[1]{\href{https://orcid.org/#1}{\includegraphics[height=\fontcharht\font`\B]{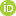}}}

\usepackage{pdfpages}
\usepackage{pgffor}
\usepackage{etoolbox}
\makeatletter
\patchcmd{\@outputpage@head}{\@ifx{\LS@rot\@undefined}{}{\LS@rot}}{}{}{}
\makeatother

\begin{document}

\title[]{Magnetic Trampoline Resonators from \ce{(La{,}Sr)MnO3} Single-Crystal Thin Films}

\author{Nicola~\surname{Manca}\,\orcidicon{0000-0002-7768-2500}}
\email{nicola.manca@spin.cnr.it}
\affiliation{CNR-SPIN, C.so F.\,M.~Perrone, 24, 16152 Genova, Italy}
\author{Dhavalkumar~\surname{Mungpara}\,\orcidicon{}}
\affiliation{Institute of Nanostructure and Solid State Physics, University of Hamburg, Germany}
\author{Leon\'elio~\surname{Cichetto}~Jr\,\orcidicon{0000-0002-5894-8852}}
\affiliation{CNR-SPIN, C.so F.\,M.~Perrone, 24, 16152 Genova, Italy}
\author{Alejandro~E.~\surname{Plaza}\,\orcidicon{0000-0002-2538-7585}}
\affiliation{CNR-SPIN, C.so F.\,M.~Perrone, 24, 16152 Genova, Italy}
\author{Gianrico~\surname{Lamura}\,\orcidicon{}}
\affiliation{CNR-SPIN, C.so F.\,M.~Perrone, 24, 16152 Genova, Italy}
\author{Alexander~\surname{Schwarz}\,\orcidicon{}}
\affiliation{Institute of Nanostructure and Solid State Physics, University of Hamburg, Germany}
\author{Daniele~\surname{Marré}\,\orcidicon{0000-0002-6230-761X}}
\affiliation{Dipartimento di Fisica, Università degli Studi di Genova, 16146 Genova, Italy}
\affiliation{CNR-SPIN, C.so F.\,M.~Perrone, 24, 16152 Genova, Italy}
\author{Luca~\surname{Pellegrino}\,\orcidicon{0000-0003-2051-4837}}
\affiliation{CNR-SPIN, C.so F.\,M.~Perrone, 24, 16152 Genova, Italy}


\begin{abstract}

    Micro-electro-mechanical resonators employing a magnetic element have been proposed for magnetic field sensing applications, but the integration of magnetic materials with standard semiconductor compounds is challenging and requires complex fabrication protocols. We present a different approach relying on \ce{(La_{0.7}{,}Sr_{0.3})MnO3} (LSMO), an oxide compound that works both as structural element for the resonator and functional magnetic layer. Suspended trampolines are realized in a single step process from LSMO thin films and show quality factor up to 60k and $f\cdot Q$ products reaching $\mathrm{10^{10}}$\,Hz. Their magnetic properties are probed by a SQUID magnetometer and magnetic force microscopy, showing saturation magnetization of 240\,kA/m at room temperature and in-plane magnetic domains with coercivity of 2.5\,mT. Being entirely made from a magnetic material, these resonators exhibit a larger magnetic interaction volume compared to other solutions, making them ideal candidates as building blocks for high-sensitivity magnetic field sensors.
 
\end{abstract}

\maketitle

\subsubsection*{Introduction}

Micro-mechanical resonators are ideal transducers for applications requiring high sensitivity \cite{Ekinci2005, Waggoner2007}, due to the precision of frequency measurements and the continuous improvement in their mechanical quality factor ($Q$)~\cite{Villanueva2014, Engelsen2024}.
Low mechanical losses are typically achieved by combining high tensile strain and complex geometries such as hierarchical structures, perimeter-mode resonators, or membranes-based phononic crystals, leading to $Q>10^9$ at room-temperature~\cite{Bereyhi2022a, Bereyhi2022, Shin2022, Hoj2024}.
On the other hand, simple cantilever or micro-bridges resonators can be easily fabricated in large arrays and are described by analytical functions, so they are often the preferential choice for many prototypical device applications~\cite{Rugar2004, Fukuma2005, Naik2009, Poggio2010, Arlett2011}. 
Suspended trampolines fall between these approaches. They consist of a central pad anchored by thin tethers to an external frame: a design that allows for high tensile strain while balancing geometric complexity and mechanical performances \cite{Romero2020}.
Trampolines resonators have been employed in a variety of sensing experiments, including displacement detectors \cite{Chien2020}, THz bolometers \cite{Vicarelli2022}, free-space opto-mechanics \cite{Manjeshwar2023}, pressure sensors \cite{Reinhardt2024}, or room-temperature quantum opto-mechanics \cite{Norte2016}.

Standard semiconductors the best choice for high-$Q$ mechanical resonators because of their reliable fabrication protocols, excellent material quality, and the integration with existing CMOS technology \cite{Ovartchaiyapong2012, Tao2014, Heritier2018, Xu2024, Manjeshwar2023}. However, the realization of high-$Q$ factor micro-bridge resonators based on transition metal oxides has been already recently reported \cite{Manca2024a}. This opens new possibilities in the realization of functional micro-mechanical devices thanks to the possibility offered by TMOs to combine materials having different physical properties into epitaxial heterostructures \cite{Zubko2011}. 

Different high-sensitivity magnetometers based on the integration of magnetic materials and mechanical resonators have been already reported in literature \cite{Forstner2012, Fischer2019, Maspero2021}, and their further implementation using TMOs might improve their performances in terms of magnetic interaction volume, stress engineering and multifunctional integrations.
A ultra-sensitive magnetic field sensor based on the coupling between a micro-mechanical magnetic sensor and a high-T$_{\mathrm{c}}$ superconductor field-to-gradient converter \cite{Pellegrino2021a} has been recently proposed. A possible implementation of the aforementioned device could be realized in a full-oxide heterostructure comprising a superconducting circuit and a mechanical resonator made of a magnetic material.
Among TMOs, \ce{(La{,}Sr)MnO3} (LSMO) shows high saturation magnetization and easy magnetization direction that can be controlled by shape anisotropy \cite{Berndt2000}. LSMO micro-bridge resonators show high tensile strain and $Q$-factor in the range of few tens of thousands \cite{Manca2022}, so this compound is a candidate for the realization of more complex TMO-based micro-mechanical magnetic sensors.

In this work, we discuss the fabrication of magnetic trampoline resonators in view of their applications in mechanical magnetometers working at room temperature or liquid nitrogen.
Mechanical properties of LSMO trampolines having total diagonal up to 700\,{\textmu}m were characterized in terms of stress and quality factor and the results compared to numerical simulations.
Magnetic properties of bare LSMO thin films are measured by SQUID magnetometry, while the magnetic microstructure of LSMO trampolines is investigated by room-temperature field-dependent magnetic force microscopy.

\subsubsection*{Device Fabrication}

\begin{figure}[]
\includegraphics[width=\linewidth]{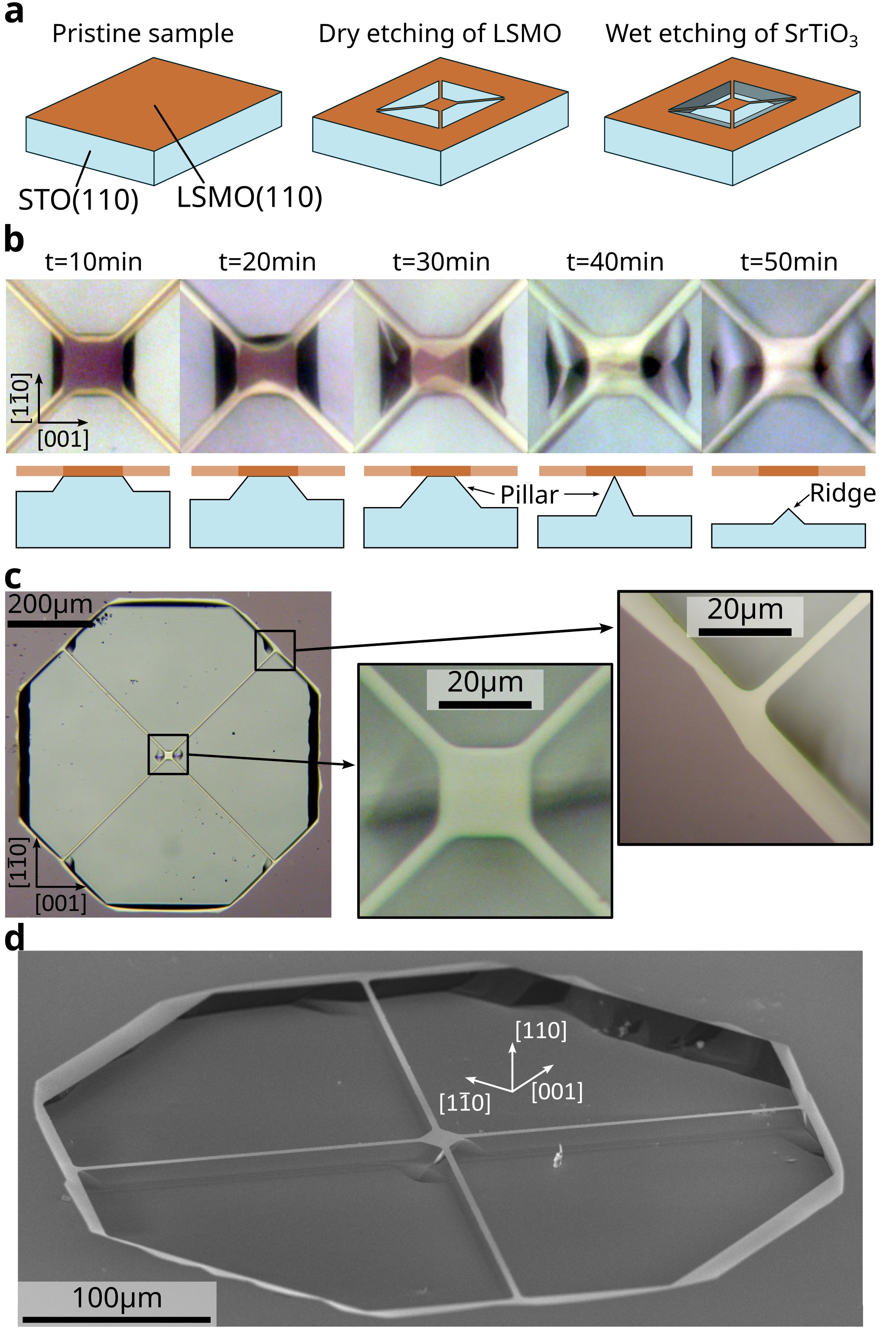}
\caption{\label{fig:tramps}
         Fabrication of LSMO trampolines.
         (a) Main steps of the trampolines fabrication process.
         (b) Optical micrographs in reflected light of the pad (20$\times$20\,{\textmu}m$^2$) of a LSMO trampoline showing the time evolution of the release process. Below each picture, we included a schematic drawing of the vertical cross-section of the device.
         (c) Optical micrograph of a 700\,{\textmu}m LSMO trampoline after the fabrication process. The pad is a 20\,{\textmu}m square and the tethers' width is 4\,\textmu m.
         (d) Scanning electron microscope image of a LSMO trampoline. 
       }
\end{figure}

Epitaxial \ce{(La{,}Sr)MnO3} thin films having thickness of 100\,nm were grown by pulsed laser deposition on top of $5\times5$\,mm$^2$  single-crystal \ce{SrTiO3(110)} substrates, as reported in the Experimental Section.
The main steps of the fabrication process of LSMO trampolines are illustrated in Figure~\ref{fig:tramps}a and follow the protocol already employed in Ref.~\onlinecite{Manca2022}. 
The trampoline geometry is transferred to the LSMO thin films by UV lithography followed by physical etching by Ar ion milling.
Trampolines are then made suspended by immersion of the sample into a 5\,\% HF water solution which corrodes the STO substrate without affecting the LSMO layer \cite{Ceriale2014}. Details of the fabrication protocol are reported in the Experimental Section.
Optical micrographs of a trampoline pad during the wet etching process are reported in Fig. \ref{fig:tramps}b.
These pictures were acquired every 10\,mins by taking the sample out of the acid bath and keeping it in deionized water and show the progression of the trampoline's release from the substrate. Clamped regions are dark/purple while the suspended LSMO film is light/yellow.
The tethers have a width of about 4\,{\textmu}m and get suspended in about 20\,min, while the pad, a square of 20\,{\textmu}m side, is fully released in less than an hour.
STO etching anisotropy determines an asymmetry in the pillar below the squared pad during the release that progressively becomes rectangular or even butterfly-like. Once the release is completed, this pillar leaves a ridge below the trampoline, which is oriented parallel to the [001] lattice direction (see the rightmost panel of Fig.~\ref{fig:tramps}b).
This is also visible in Fig.~\ref{fig:tramps}c, showing an optical micrograph of a LSMO trampoline having total diagonal of 700\,{\textmu}m after being removed from the acid bath and dried in a critical-point dryer system.
Here, the surface of the pit in-between the tethers is flat and maintains its original [110] orientation, with no faceting, because of the STO etching anisotropy \cite{Plaza2021}.
The black regions at the edges of the trampoline pit are flat surfaces that do no reflect lights back into the microscope objective and are due to the faceting of the STO during the etching. These regions are also visible in the scanning electron microscope image reported in Fig.~\ref{fig:tramps}d.

\begin{figure*}[]
\includegraphics[width=\linewidth]{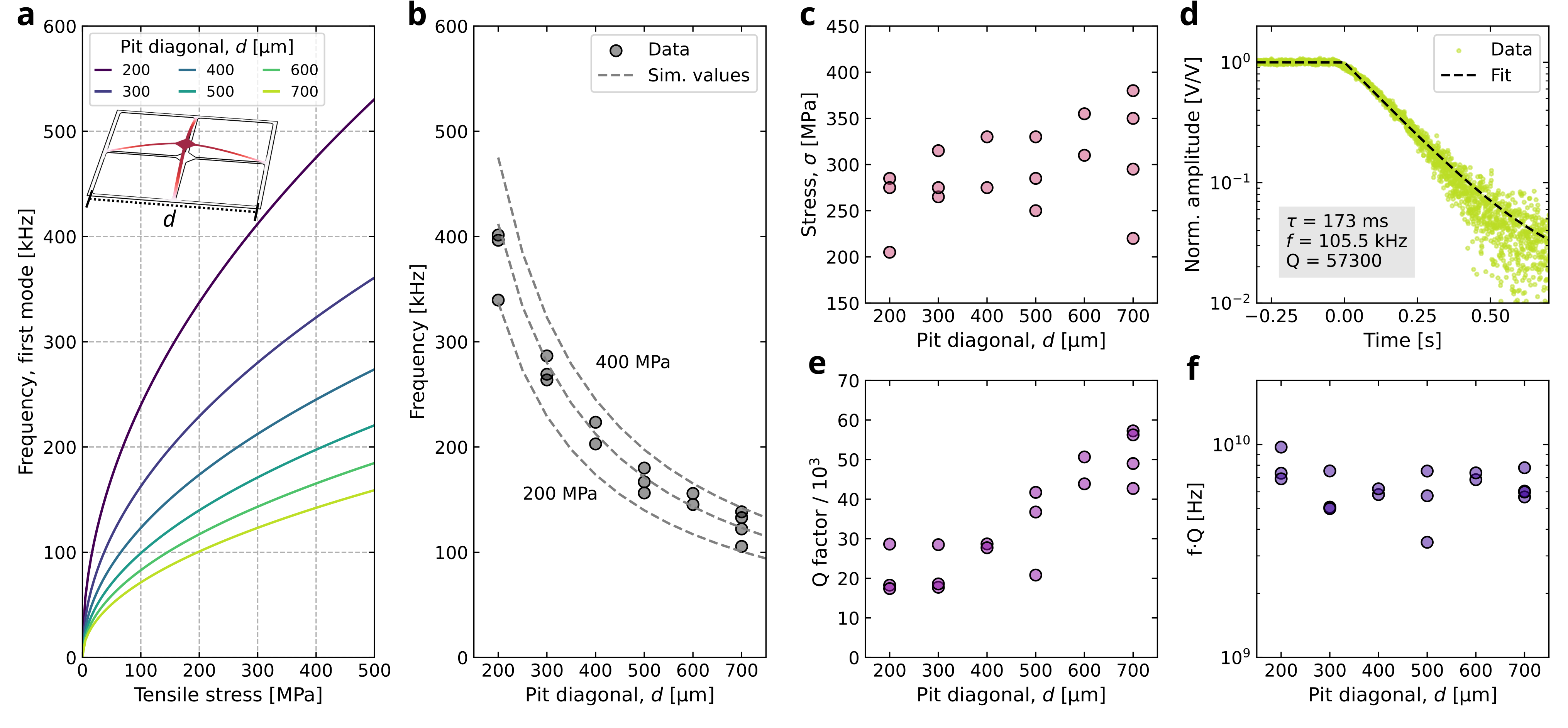}
\caption{\label{fig:mec}
Mechanical characterization of LSMO trampolines.
(a) Stress an size dependence of the first mode resonance frequency for LSMO trampolines calculated by finite element analysis.
(b) Resonance frequencies of the first mode of of LSMO trampolines. Dashed lines indicates the stress calculated from (a) and are spaced 100\,MPa.
(c) Isotropic in-plane tensile stress of the trampolines obtained by comparing the simulation results reported in (a) and the experimental data reported in (b). 
(d) Ring-down measurement of the first mode of the LSMO trampoline showing the highest measured $Q$-factor, having $d$=700\,{\textmu}m.
(e) $Q$-factor and corresponding $f \cdot Q$ products (f) as a function of the pit diagonal of the trampolines reported in (b).
}
\end{figure*}

In-plane etching anisotropy of STO(110) is crucial in determining the release process and should be taken into account when designing the tethers geometry.
This is because etching rates and walls profiles along the [001] and [1\={1}0] directions are different and tethers aligned along these two axis would have different suspension time and clamping conditions~\cite{Manca2024}.
By designing the tethers rotated 45° in-plane, we make them all suspended at the same time and with the same clamping points geometry (except for mirror symmetry).
Anisotropic under-etching is better visible when looking at the perimeter of the trampoline pit.
The edges parallel to the [001] direction (the top/bottom ones in Fig.~\ref{fig:tramps}c) show about 20\,{\textmu}m of under-etching, while for those parallel to the [001] direction (the left/right ones in Fig.~\ref{fig:tramps}c) the under-etching is just few micrometers. The edges parallel to the four [1\={1}1] equivalent directions show under-etching of about 10\,{\textmu}m, a value in-between the two previous cases.
Different clamping geometry could be obtained by designing the clamping points to be along edges having different alignment, as an example by choosing the direction where the under-etching is lower.
However, a lower symmetry in the tether configuration may result in a deformed structure upon release due to stress relaxation.
The interplay between the geometry of resonators having more than two clamping points and their mechanical properties is of great interest for ultra-high quality factor Si-based resonators~\cite{Shin2022, Bereyhi2022a} and a future research direction for full-oxide mechanical devices as well.

\subsubsection*{Mechanical characterization}

Mechanical properties of LSMO trampolines were probed by measuring the resonance frequency and quality factor of devices having different size. Details of the experimental setup and measurement protocol are reported in the Experimental Section.
The characteristic length for the trampolines is the distance between two opposite tether clamping points, i.e.\ the pit diagonal.
Since no analytical solution exists for the modes frequencies of a trampoline resonator, we performed a numerical simulation by finite element analysis to calculate the resonance frequencies as a function of the stress ($\sigma$) for different pit diagonal ($d$) values. Details of the simulation are reported in the Experimental Section.
Results for the frequencies of the first mode are shown in Figure~\ref{fig:mec}a, where the panel inset shows the corresponding mode shape.  These data are employed to evaluate the stress of the trampolines from the frequency measurements as discussed below.
Fig.~\ref{fig:mec}b, shows the first-mode resonance frequency measured on a set of LSMO trampolines. Here, the gray dashed lines are constant-stress curves in the frequency--length space parameter. They are spaced 100\,MPa apart and correspond to vertical cuts of Fig.~\ref{fig:mec}a.
By comparing the measured frequencies with the numerical simulations we obtain the corresponding in-plane stress reported in Fig.~\ref{fig:mec}c.
To do so, for each measured device we look for the stress value of Fig.~\ref{fig:mec}a providing the closest-matching frequency. These numerical solutions were calculated with a step size of 5\,MPa, that allows to to evaluate the stress of the trampolines with reasonable precision.
Although the limited number of measured devices does not allows for a robust statistical analysis, the average stress magnitude is about 300\,MPa.
These values are in good agreement with previous measurements performed on LSMO micro-bridges, although here few devices show stress reaching 400\,MPa, which is the highest value reported so far for oxide resonators \cite{Manca2022, Manca2024}.
The origin of such high value could be related to slightly different local growth conditions, such as temperature gradients or alignment of the sample with respect to the plasma plume, suggesting that tensile stress may be improved by tuning the growth conditions.
We measured the quality factor of the first mode of the trampolines from ring-down measurements. The experimental data of the trampoline having the highest $Q$-factor is shown in Fig.~\ref{fig:mec}d, corresponding to a $d$=700\,{\textmu}m device.
Fig.~\ref{fig:mec}e shows the $Q$-factor of all the resonator reported in Fig.~\ref{fig:mec}a.
They shows a steady increase of the $Q$-factor starting from 400\,{\textmu}m of diagonal and getting close to 60k for 700\,{\textmu}m ones. The linear increase of the $Q$-factor with size observed for $d>$400\,{\textmu}m, indicates that the mechanical losses are dominated by dissipation-dilution mechanism, suggesting that LSMO intrinsic $Q$-factor could be obtained by realizing micro-bridge resonators in this length range \cite{Engelsen2024}.
From these data it is possible to evaluate the $f \cdot Q$ product of LSMO trampolines, that are reported in Fig~\ref{fig:mec}f. For all the measured trampolines these are just below $10^{10}$\,Hz and show no size dependence.

\subsubsection*{Magnetic Characterization}

\begin{figure}[]
  \includegraphics[width=\linewidth]{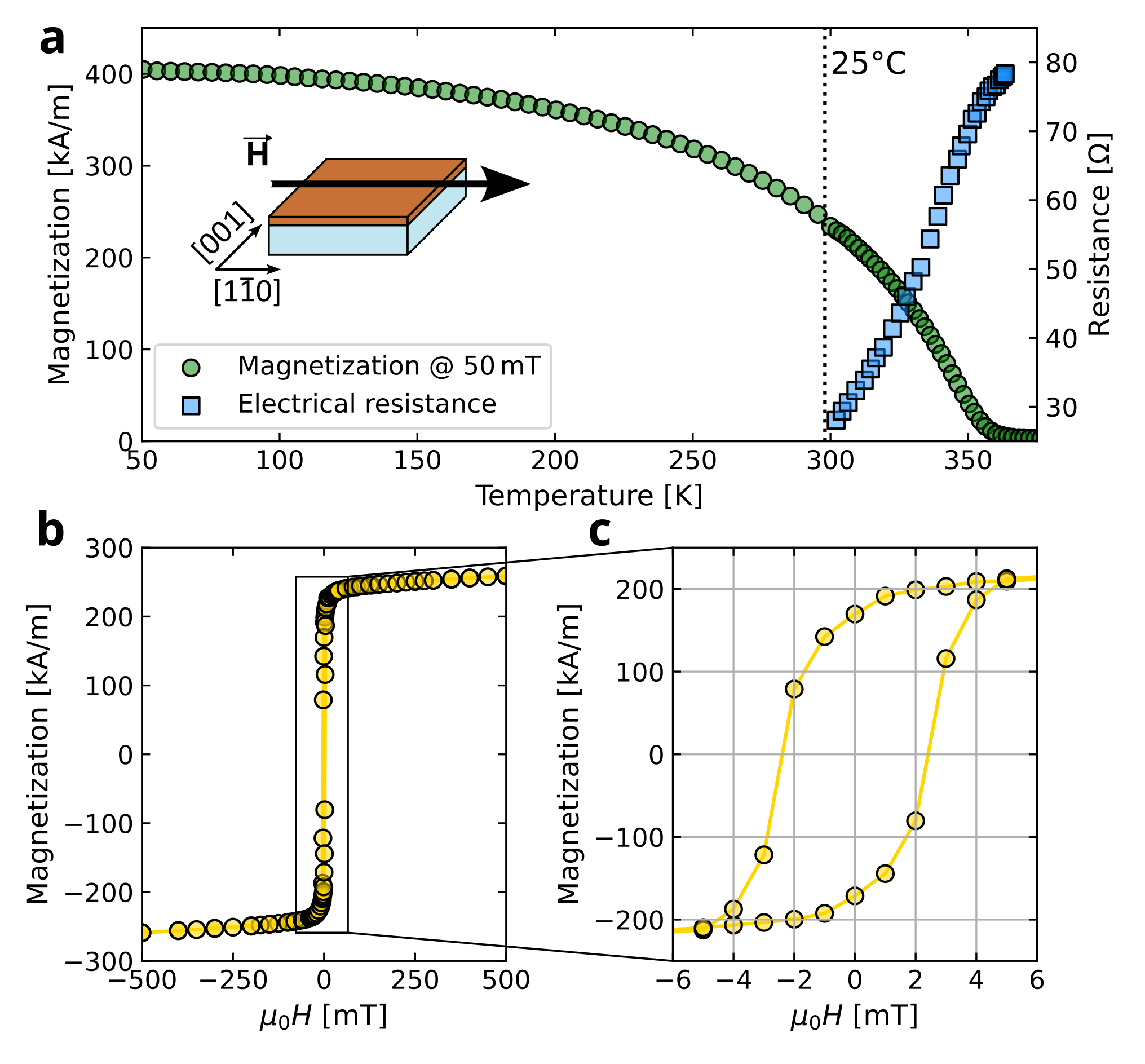}
\caption{\label{fig:mag}
Magnetic characterization of LSMO(110) thin films.
(a) Temperature dependence of the in-plane magnetization along the [1\={1}0] lattice direction (green dots) and normalized electrical resistance measured in four probe configuration (blue squares).
(b) Magnetization loop at 25\,{\textcelsius} measured in the $\pm$500\,mT range.
(c) Magnification of the data reported in (b) to better show the magnetic hysteresis.
}
\end{figure}

The magnetic properties of a pristine LSMO thin film grown on top of STO(110) were measured by a commercial dc-SQUID magnetometer (MPMS2 by Quantum Design, Inc. USA).
Figure~\ref{fig:mag}a shows the temperature dependence of the field-cooled magnetization $M(T)$ with $\mu_0H$=50\,mT applied along the [1\={1}0] direction, as indicated in the inset. This is the "easy magnetization axis" (see Supplementary Materials Sec.~I for further details).
The Curie temperature ($T_\mathrm{C}$) has been defined by the maximum of $\partial M / \partial T$ and it results $T_\mathrm{C}$=344$\pm$2\,K as expected for \ce{(La_{0.7}{,}Sr_{0.3})MnO3} \cite{Tokura1999}, and the measured magnetization reaches 400\,kA/m at 50\,K.
The onset of the magnetic transition can be also indirectly identified by measuring the temperature dependence of the electrical resistivity $R(T)$, which shows a point of inflection just below the Curie temperature \cite{Mitchell1996}. In the Supplementary Material Sec. II, we compare the R(T) characteristics of the LSMO films employed for the different experiments discussed in this work, all showing a similar $R(T)$ behavior.
The magnetization loop measured at room-temperature (25\,{\textcelsius}) is reported in Fig.~\ref{fig:mag}b and c. Saturation magnetization is above 250\,kA/m, with about 170\,kA/m of remanence and about 2.5\,mT of coercitivity.

\begin{figure}[b]
\includegraphics[width=\linewidth]{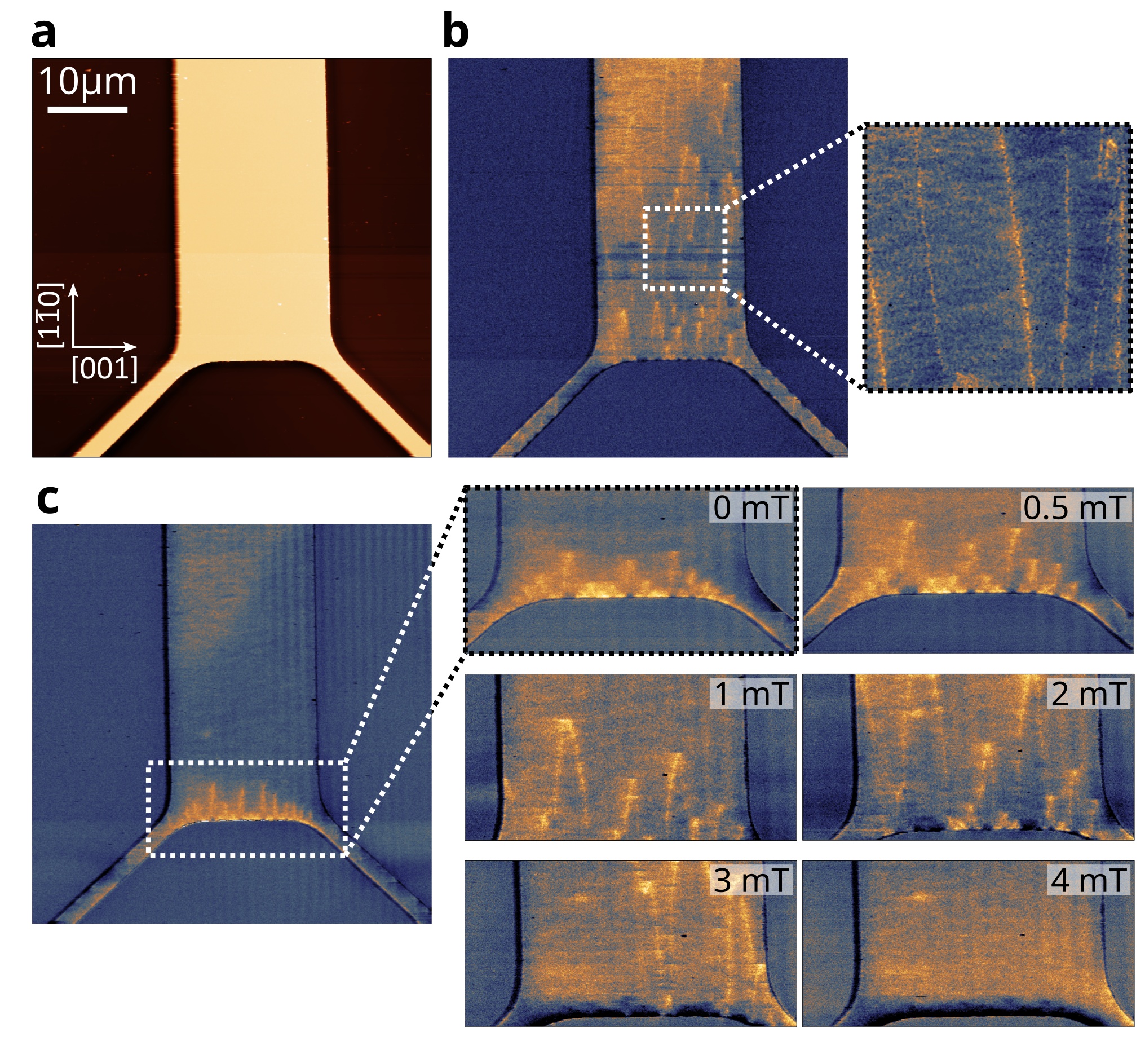}
\caption{\label{fig:mfm}
	 Micro-magnetic characterization of LSMO(110) clamped trampolines.
         (a) Surface topography of a LSMO rectangular trampoline measured by atomic force microscopy using a magnetic tip.
         (b) Magnetic phase contrast of the region shown in (a) without pre-magnetization. Edges of magnetic domains are visible as lighter vertical stripes.
         (c) Magnetic microstructure after applying 20\,mT in-plane. Magnified regions shows the domains evolution while applying a magnetic field in the reverse direction.
	}
\end{figure}

For our envisaged application, i.e. using TMO based trampoline-shaped resonators as magnetic field sensors, we are also interested in their magnetic microstructure, which we investigated by magnetic force microscopy (MFM). Details of our specific MFM set-up are described in the Experimental Section. 
For this experiment we prepared a 100\,nm tick trampoline-shaped LSMO structure in a 240\,{\textmu}m $\times$ 280\,{\textmu}m frame with a 20$\times$60\,{\textmu}m$^2$ rectangular central pad and four  100\,{\textmu}m-long tethers.
For the selected dimensions, shape anisotropy prefers an in-plane magnetization (typical for thin films) oriented along the long axis of the rectangle, i.e., the [1\={1}0] direction, which is also the easy axis of magnetization.
This trampoline was not free-hanging, but was still attached to its growth substrate, i.e.\ MFM measurements were performed after dry etching the LSMO and the following cleaning of the surface. This was necessary because in our set-up the interaction between the MFM tip and free-hanging trampolines made MFM data acquisition unreliable, often resulting in the destruction of the device. However, since the trampoline geometry retains its tensile stress, we do not expect a relevant difference in the micro-magnetic configuration between the released and unreleased configuration.

The surface topography of the trampoline is reported in Figure~\ref{fig:mfm}a, showing almost all of the central pad and parts of the two bottom tethers, while the frame is not visible in this image.
Fig.~\ref{fig:mfm}b shows the pristine magnetic contrast in zero field. In the magnified inset, bright lines predominantly aligned along the long axis of the rectangle can be identified. In between these lines the contrast is rather homogeneous. Note, that a line-like contrast is also visible on the tethers. However, here we are mainly concerned with the domains structure on the central pad.
Due to their geometry, MFM tips are mostly sensitive to the out-of-plane component of the stray field emanating from the sample. Thus, line-like features in MFM images of magnetic thin films indicate a domain-wall contrast and thus an in-plane magnetization of the sample. Moreover, a simple dark (or bright) line indicates the presence of Bloch-type domain walls, in which the magnetization in the domain wall rotates out-of-plane. 
We conclude that the observed MFM contrast is consistent with an in-plane magnetization with the easy axis of magnetization oriented along the [1\={1}0] direction. 
Fig.~\ref{fig:mfm}c shows the remanent magnetic structure in zero field, i.e., after saturating the whole LSMO film in an in-plane external field of 20\,mT applied along the long axis of the rectangle (actually, saturation is already observed at about 4 mT). In saturation, the contrast is homogeneous across the trampoline (its in a single domain state), with a dark and bright contrast at the opposite short sides of the central pad, where the magnetic field lines leave the sample pointing in and out of the sample plane. The first domains with an in-plan magnetization antiparallel to the direction of the saturating field start to nucleate as spike domains at the short edges of the rectangle. In remanence they are only a few {\textmu}m long and cover less than 3\,\% of the area of the central pad. Upon reversing the direction of the external magnetic field, the domains grow further until the central pad is saturated in opposite direction. This growth is displayed in the image series in Fig~\ref{fig:mfm}c.
Easy axis of magnetization, saturation magnetization, domain nucleation and growth visible in the MFM data is in good agreement with the measured $M(H)$ hysteresis loop displayed in Fig. ~\ref{fig:mag}b, c.


\section*{Conclusion}

In conclusion, suspended magnetic trampolines having diagonal length up to 700\,{\textmu}m were realized from LSMO epitaxial thin films grown on top of \ce{SrTiO3}(110) substrates. These resonators have in-plane tensile stress typically in between 250\,MPa and 350\,MPa and mechanical quality factor up to 60k with signature of dissipation-dilution dominated mechanical losses. Room-temperature magnetic characterization of the bare film before the micro-fabrication and of the patterned trampoline before being suspended show similar coercitivity of about 2.5\,mT. Saturation magnetization at room temperature is 250\,kA/m, reaching 400\,kA/m at liquid nitrogen temperature. The combination of high mechanical quality factor and magnetic properties makes LSMO trampolines ideal candidates to realize MEMS magnetic sensors, in particular if combined with other complex oxides in multifunctional heterostructured devices.

\section*{Experimental}
\paragraph*{\ce{(La{,}Sr)MnO3} thin films deposition:}
LSMO thin films are grown by pulsed laser deposition on top of not-terminated 5$\times$5\,mm$^2$ single-crystal \ce{SrTiO3(110)} substrates.
The substrate is mounted in a stainless steel sample-holder that is kept facing downwards during the deposition.
The growth chamber has a base pressure below $1 \cdot 10^{-8}$\,mbar, while during the growth the oxygen pressure is $1 \cdot 10^{-4}$\,mbar.
The KrF excimer laser (248 nm) energy density is 0.6\,J\,cm${}^{-2}$, the laser repetition rate is 2\,Hz, and the target--sample distance is about 45\,mm.
During the growth, the temperature is monitored by a infrared pyrometer measuring the backside of 
the sample holder.
LSMO thin films emplyed in this work were grown at 850\,{\textcelsius}.  
The growth rate was 0.56\,nm/min, as measured by monitoring the intensity oscillations of reflection high-energy electron diffraction signal.

\paragraph*{Microfabrication:}
LSMO trampolines are realized by UV photolithography.
The SPR-220-4.5 photo-resist from Microposit\texttrademark is deposited by drop casting and spinning speed of 6000\,RPM for 45\,s.
Resist backing is performed at 125\,{\textcelsius} for 150\,s.
Exposed LSMO regions are etched in about 45\,mins by \ce{Ar^+} ion milling having energy of 500\,eV and current density of 0.2\,mA/cm$^2$.
The trampolines are then made suspended by immersion in 5\,\% HF in water solution at 35\,{\textcelsius} for about one hour. During the corrosion step the sample is kept above a magnetic stirrer spinning at about 200\,RPM.
At the end of the chemical etching step, the sample is immersed for about 30\,mins in 5\,\% \ce{H3PO4} in water solution to mechanically remove insoluble residues of the \ce{SrTiO3} corrosion in HF. This solution is kept at 40\,{\textcelsius} and magnetically stirred at 100\,RPM.
When ready, the sample is cleaned by immersion firstly in water and then in ethanol baths.
Finally it is dried in a \ce{L-CO2} critical point drier.

\paragraph*{Mechanical measurements:}
Mechanical measurements were performed in a custom setup featuring active temperature control by Peltier electro-thermal module.
Samples were kept at were 25\,{\textcelsius} and residual air pressure was below 1$\cdot$10$^{-4}$\,mbar.
Device motion was probed in the optical lever detection scheme by a four-quadrant photo-diode.
Mechanical excitation was provided by an AC-biased piezoelectric element glued by ceramic epoxy near to the device.

\paragraph*{Finite element analysis:}
Numerical simulation of the trampolines resonances were performed in \textsc{COMSOL} Multiphysics{\textregistered} v.6.2 using the membrane module.
Model geometry was a 20\,{\textmu}m$^2$ squared pad with 5\,{\textmu}m tethers clamped to an external squared frame. Tethers' length and frame size varied following the $d$ parameter. Frame had ``Fixed'' condition, while the rest of the edges were set as ``Free boundary''. 
Membrane thickness was set to 100\,nm.
Mesh was generated by ``Physics-Controlled mesh'' sequence type with ``Extremely fine'' element size setting.
Isotropic stress was imposed all over the membrane as initial condition and let to relax in the suspended regions.
The study type was ``pre-stressed eigenfrequency'' with geometric nonlinearity enabled and parametric sweep over stress value and pit diameter.
The simulation file is included in the dataset (see ``Open Data'').

\paragraph*{Magnetic force microscopy (MFM) measurements:}
MFM was employed to probe the magnetic microstructure of an LSMO thin film with an unreleased trampoline geometry at room temperature. We used  an MFM cantilever (resonance frequency: $\sim$75\,kHz; spring constant: $\sim$3\,N/m) with a hard magnetic coating and out-of-plane sensitivity. We used a Nanoscope IIIa instrument with a homemade calibrated dipole electromagnet (two copper coils in a Helmholtz configuration with permalloy yokes), which enables us to generate an in-plane magnetic field of about $\pm$20\,mT 
at the sample position. For imaging we used the so-called lift mode with amplitude modulation (tapping mode). This mode of operation generates a topography image and an MFM image in one go. Both images are recorded line-by-line. First, the surface profile, i.e. the topography $z(x,y)$, is recorded with an active $z$-piezo feedback using the amplitude signal. Thereafter, the same line is recorded without feedback at a constant lift height of 80\,nm above the surface by using the previously recorded topography data. The recorded variation of the phase signal $\phi(x,y)$ is induced by the long-range magnetostatic tip-sample interaction, which reflects the magnetic microstructure of the sample. To minimize long-range electrostatic contributions to the phase signal, we compensated the contact-potential between the magnetic tip and the LSMO surface by applying a bias voltage of \textminus 1.8\,V. 

\section*{Acknowledgements}

We thank Cristina Bernini for the support with the scanning electron microscope.
This work was carried out under the OXiNEMS project (\href{www.oxinems.eu}{www.oxinems.eu}). This project has received funding from the European Union’s Horizon 2020 research and innovation programme under Grant Agreement No. 828784.

\section*{Open Data}

The numerical data shown in figures of the manuscript
can be downloaded from the Zenodo online repository:
~\href{http://dx.doi.org/10.5281/zenodo.14711494}{http://dx.doi.org/10.5281/zenodo.14711494}

\bibliography{Library.bib}

\begin{thebibliography}{36}%
\makeatletter
\providecommand \@ifxundefined [1]{%
 \@ifx{#1\undefined}
}%
\providecommand \@ifnum [1]{%
 \ifnum #1\expandafter \@firstoftwo
 \else \expandafter \@secondoftwo
 \fi
}%
\providecommand \@ifx [1]{%
 \ifx #1\expandafter \@firstoftwo
 \else \expandafter \@secondoftwo
 \fi
}%
\providecommand \natexlab [1]{#1}%
\providecommand \enquote  [1]{``#1''}%
\providecommand \bibnamefont  [1]{#1}%
\providecommand \bibfnamefont [1]{#1}%
\providecommand \citenamefont [1]{#1}%
\providecommand \href@noop [0]{\@secondoftwo}%
\providecommand \href [0]{\begingroup \@sanitize@url \@href}%
\providecommand \@href[1]{\@@startlink{#1}\@@href}%
\providecommand \@@href[1]{\endgroup#1\@@endlink}%
\providecommand \@sanitize@url [0]{\catcode `\\12\catcode `\$12\catcode
  `\&12\catcode `\#12\catcode `\^12\catcode `\_12\catcode `\%12\relax}%
\providecommand \@@startlink[1]{}%
\providecommand \@@endlink[0]{}%
\providecommand \url  [0]{\begingroup\@sanitize@url \@url }%
\providecommand \@url [1]{\endgroup\@href {#1}{\urlprefix }}%
\providecommand \urlprefix  [0]{URL }%
\providecommand \Eprint [0]{\href }%
\providecommand \doibase [0]{https://doi.org/}%
\providecommand \selectlanguage [0]{\@gobble}%
\providecommand \bibinfo  [0]{\@secondoftwo}%
\providecommand \bibfield  [0]{\@secondoftwo}%
\providecommand \translation [1]{[#1]}%
\providecommand \BibitemOpen [0]{}%
\providecommand \bibitemStop [0]{}%
\providecommand \bibitemNoStop [0]{.\EOS\space}%
\providecommand \EOS [0]{\spacefactor3000\relax}%
\providecommand \BibitemShut  [1]{\csname bibitem#1\endcsname}%
\let\auto@bib@innerbib\@empty
\bibitem [{\citenamefont {Ekinci}\ and\ \citenamefont
  {Roukes}(2005)}]{Ekinci2005}%
  \BibitemOpen
  \bibfield  {author} {\bibinfo {author} {\bibfnamefont {K.~L.}\ \bibnamefont
  {Ekinci}}\ and\ \bibinfo {author} {\bibfnamefont {M.~L.}\ \bibnamefont
  {Roukes}},\ }\bibfield  {title} {\bibinfo {title} {Nanoelectromechanical
  systems},\ }\href {https://doi.org/10.1063/1.1927327} {\bibfield  {journal}
  {\bibinfo  {journal} {Review of Scientific Instruments}\ }\textbf {\bibinfo
  {volume} {76}},\ \bibinfo {pages} {061101} (\bibinfo {year}
  {2005})}\BibitemShut {NoStop}%
\bibitem [{\citenamefont {Waggoner}\ and\ \citenamefont
  {Craighead}(2007)}]{Waggoner2007}%
  \BibitemOpen
  \bibfield  {author} {\bibinfo {author} {\bibfnamefont {P.~S.}\ \bibnamefont
  {Waggoner}}\ and\ \bibinfo {author} {\bibfnamefont {H.~G.}\ \bibnamefont
  {Craighead}},\ }\bibfield  {title} {\bibinfo {title} {Micro- and
  nanomechanical sensors for environmental, chemical, and biological
  detection},\ }\href {https://doi.org/10.1039/b707401h} {\bibfield  {journal}
  {\bibinfo  {journal} {Lab Chip}\ }\textbf {\bibinfo {volume} {7}},\ \bibinfo
  {pages} {1238} (\bibinfo {year} {2007})}\BibitemShut {NoStop}%
\bibitem [{\citenamefont {Villanueva}\ and\ \citenamefont
  {Schmid}(2014)}]{Villanueva2014}%
  \BibitemOpen
  \bibfield  {author} {\bibinfo {author} {\bibfnamefont {L.~G.}\ \bibnamefont
  {Villanueva}}\ and\ \bibinfo {author} {\bibfnamefont {S.}~\bibnamefont
  {Schmid}},\ }\bibfield  {title} {\bibinfo {title} {Evidence of {{Surface
  Loss}} as {{Ubiquitous Limiting Damping Mechanism}} in {{SiN Micro-}} and
  {{Nanomechanical Resonators}}},\ }\href
  {https://doi.org/10.1103/PhysRevLett.113.227201} {\bibfield  {journal}
  {\bibinfo  {journal} {Phys. Rev. Lett.}\ }\textbf {\bibinfo {volume} {113}},\
  \bibinfo {pages} {227201} (\bibinfo {year} {2014})}\BibitemShut {NoStop}%
\bibitem [{\citenamefont {Engelsen}\ \emph {et~al.}(2024)\citenamefont
  {Engelsen}, \citenamefont {Beccari},\ and\ \citenamefont
  {Kippenberg}}]{Engelsen2024}%
  \BibitemOpen
  \bibfield  {author} {\bibinfo {author} {\bibfnamefont {N.~J.}\ \bibnamefont
  {Engelsen}}, \bibinfo {author} {\bibfnamefont {A.}~\bibnamefont {Beccari}},\
  and\ \bibinfo {author} {\bibfnamefont {T.~J.}\ \bibnamefont {Kippenberg}},\
  }\bibfield  {title} {\bibinfo {title} {Ultrahigh-quality-factor micro- and
  nanomechanical resonators using dissipation dilution},\ }\href
  {https://doi.org/10.1038/s41565-023-01597-8} {\bibfield  {journal} {\bibinfo
  {journal} {Nat. Nanotechnol.}\ ,\ \bibinfo {pages} {1}} (\bibinfo {year}
  {2024})}\BibitemShut {NoStop}%
\bibitem [{\citenamefont {Bereyhi}\ \emph
  {et~al.}(2022{\natexlab{a}})\citenamefont {Bereyhi}, \citenamefont {Beccari},
  \citenamefont {Groth}, \citenamefont {Fedorov}, \citenamefont {Arabmoheghi},
  \citenamefont {Kippenberg},\ and\ \citenamefont {Engelsen}}]{Bereyhi2022a}%
  \BibitemOpen
  \bibfield  {author} {\bibinfo {author} {\bibfnamefont {M.~J.}\ \bibnamefont
  {Bereyhi}}, \bibinfo {author} {\bibfnamefont {A.}~\bibnamefont {Beccari}},
  \bibinfo {author} {\bibfnamefont {R.}~\bibnamefont {Groth}}, \bibinfo
  {author} {\bibfnamefont {S.~A.}\ \bibnamefont {Fedorov}}, \bibinfo {author}
  {\bibfnamefont {A.}~\bibnamefont {Arabmoheghi}}, \bibinfo {author}
  {\bibfnamefont {T.~J.}\ \bibnamefont {Kippenberg}},\ and\ \bibinfo {author}
  {\bibfnamefont {N.~J.}\ \bibnamefont {Engelsen}},\ }\bibfield  {title}
  {\bibinfo {title} {Hierarchical tensile structures with ultralow mechanical
  dissipation},\ }\href {https://doi.org/10.1038/s41467-022-30586-z} {\bibfield
   {journal} {\bibinfo  {journal} {Nat Commun}\ }\textbf {\bibinfo {volume}
  {13}},\ \bibinfo {pages} {3097} (\bibinfo {year} {2022}{\natexlab{a}})},\
  \Eprint {https://arxiv.org/abs/2103.09785} {arXiv:2103.09785} \BibitemShut
  {NoStop}%
\bibitem [{\citenamefont {Bereyhi}\ \emph
  {et~al.}(2022{\natexlab{b}})\citenamefont {Bereyhi}, \citenamefont
  {Arabmoheghi}, \citenamefont {Beccari}, \citenamefont {Fedorov},
  \citenamefont {Huang}, \citenamefont {Kippenberg},\ and\ \citenamefont
  {Engelsen}}]{Bereyhi2022}%
  \BibitemOpen
  \bibfield  {author} {\bibinfo {author} {\bibfnamefont {M.~J.}\ \bibnamefont
  {Bereyhi}}, \bibinfo {author} {\bibfnamefont {A.}~\bibnamefont
  {Arabmoheghi}}, \bibinfo {author} {\bibfnamefont {A.}~\bibnamefont
  {Beccari}}, \bibinfo {author} {\bibfnamefont {S.~A.}\ \bibnamefont
  {Fedorov}}, \bibinfo {author} {\bibfnamefont {G.}~\bibnamefont {Huang}},
  \bibinfo {author} {\bibfnamefont {T.~J.}\ \bibnamefont {Kippenberg}},\ and\
  \bibinfo {author} {\bibfnamefont {N.~J.}\ \bibnamefont {Engelsen}},\
  }\bibfield  {title} {\bibinfo {title} {Perimeter {{Modes}} of
  {{Nanomechanical Resonators Exhibit Quality Factors Exceeding}}
  10{\textsubscript{9 at }}{{{\textsubscript{Room Temperature}}}}},\ }\href
  {https://doi.org/10.1103/PhysRevX.12.021036} {\bibfield  {journal} {\bibinfo
  {journal} {Phys. Rev. X}\ }\textbf {\bibinfo {volume} {12}},\ \bibinfo
  {pages} {021036} (\bibinfo {year} {2022}{\natexlab{b}})}\BibitemShut
  {NoStop}%
\bibitem [{\citenamefont {Shin}\ \emph {et~al.}(2022)\citenamefont {Shin},
  \citenamefont {Cupertino}, \citenamefont {{de Jong}}, \citenamefont
  {Steeneken}, \citenamefont {Bessa},\ and\ \citenamefont {Norte}}]{Shin2022}%
  \BibitemOpen
  \bibfield  {author} {\bibinfo {author} {\bibfnamefont {D.}~\bibnamefont
  {Shin}}, \bibinfo {author} {\bibfnamefont {A.}~\bibnamefont {Cupertino}},
  \bibinfo {author} {\bibfnamefont {M.~H.~J.}\ \bibnamefont {{de Jong}}},
  \bibinfo {author} {\bibfnamefont {P.~G.}\ \bibnamefont {Steeneken}}, \bibinfo
  {author} {\bibfnamefont {M.~A.}\ \bibnamefont {Bessa}},\ and\ \bibinfo
  {author} {\bibfnamefont {R.~A.}\ \bibnamefont {Norte}},\ }\bibfield  {title}
  {\bibinfo {title} {Spiderweb {{Nanomechanical Resonators}} via {{Bayesian
  Optimization}}: {{Inspired}} by {{Nature}} and {{Guided}} by {{Machine
  Learning}}},\ }\href {https://doi.org/10.1002/adma.202106248} {\bibfield
  {journal} {\bibinfo  {journal} {Advanced Materials}\ }\textbf {\bibinfo
  {volume} {34}},\ \bibinfo {pages} {2106248} (\bibinfo {year}
  {2022})}\BibitemShut {NoStop}%
\bibitem [{\citenamefont {H{\o}j}\ \emph {et~al.}(2024)\citenamefont {H{\o}j},
  \citenamefont {Hoff},\ and\ \citenamefont {Andersen}}]{Hoj2024}%
  \BibitemOpen
  \bibfield  {author} {\bibinfo {author} {\bibfnamefont {D.}~\bibnamefont
  {H{\o}j}}, \bibinfo {author} {\bibfnamefont {U.~B.}\ \bibnamefont {Hoff}},\
  and\ \bibinfo {author} {\bibfnamefont {U.~L.}\ \bibnamefont {Andersen}},\
  }\bibfield  {title} {\bibinfo {title} {Ultracoherent {{Nanomechanical
  Resonators Based}} on {{Density Phononic Crystal Engineering}}},\ }\href
  {https://doi.org/10.1103/PhysRevX.14.011039} {\bibfield  {journal} {\bibinfo
  {journal} {Phys. Rev. X}\ }\textbf {\bibinfo {volume} {14}},\ \bibinfo
  {pages} {011039} (\bibinfo {year} {2024})}\BibitemShut {NoStop}%
\bibitem [{\citenamefont {Rugar}\ \emph {et~al.}(2004)\citenamefont {Rugar},
  \citenamefont {Budakian}, \citenamefont {Mamin},\ and\ \citenamefont
  {Chui}}]{Rugar2004}%
  \BibitemOpen
  \bibfield  {author} {\bibinfo {author} {\bibfnamefont {D.}~\bibnamefont
  {Rugar}}, \bibinfo {author} {\bibfnamefont {R.}~\bibnamefont {Budakian}},
  \bibinfo {author} {\bibfnamefont {H.~J.}\ \bibnamefont {Mamin}},\ and\
  \bibinfo {author} {\bibfnamefont {B.~W.}\ \bibnamefont {Chui}},\ }\bibfield
  {title} {\bibinfo {title} {Single spin detection by magnetic resonance force
  microscopy},\ }\href {https://doi.org/10.1038/nature02658} {\bibfield
  {journal} {\bibinfo  {journal} {Nature}\ }\textbf {\bibinfo {volume} {430}},\
  \bibinfo {pages} {329} (\bibinfo {year} {2004})}\BibitemShut {NoStop}%
\bibitem [{\citenamefont {Fukuma}\ \emph {et~al.}(2005)\citenamefont {Fukuma},
  \citenamefont {Kimura}, \citenamefont {Kobayashi}, \citenamefont
  {Matsushige},\ and\ \citenamefont {Yamada}}]{Fukuma2005}%
  \BibitemOpen
  \bibfield  {author} {\bibinfo {author} {\bibfnamefont {T.}~\bibnamefont
  {Fukuma}}, \bibinfo {author} {\bibfnamefont {M.}~\bibnamefont {Kimura}},
  \bibinfo {author} {\bibfnamefont {K.}~\bibnamefont {Kobayashi}}, \bibinfo
  {author} {\bibfnamefont {K.}~\bibnamefont {Matsushige}},\ and\ \bibinfo
  {author} {\bibfnamefont {H.}~\bibnamefont {Yamada}},\ }\bibfield  {title}
  {\bibinfo {title} {Development of low noise cantilever deflection sensor for
  multienvironment frequency-modulation atomic force microscopy},\ }\href
  {https://doi.org/10.1063/1.1896938} {\bibfield  {journal} {\bibinfo
  {journal} {Review of Scientific Instruments}\ }\textbf {\bibinfo {volume}
  {76}},\ \bibinfo {pages} {053704} (\bibinfo {year} {2005})}\BibitemShut
  {NoStop}%
\bibitem [{\citenamefont {Naik}\ \emph {et~al.}(2009)\citenamefont {Naik},
  \citenamefont {Hanay}, \citenamefont {Hiebert}, \citenamefont {Feng},\ and\
  \citenamefont {Roukes}}]{Naik2009}%
  \BibitemOpen
  \bibfield  {author} {\bibinfo {author} {\bibfnamefont {A.~K.}\ \bibnamefont
  {Naik}}, \bibinfo {author} {\bibfnamefont {M.~S.}\ \bibnamefont {Hanay}},
  \bibinfo {author} {\bibfnamefont {W.~K.}\ \bibnamefont {Hiebert}}, \bibinfo
  {author} {\bibfnamefont {X.~L.}\ \bibnamefont {Feng}},\ and\ \bibinfo
  {author} {\bibfnamefont {M.~L.}\ \bibnamefont {Roukes}},\ }\bibfield  {title}
  {\bibinfo {title} {Towards single-molecule nanomechanical mass
  spectrometry},\ }\href {https://doi.org/10.1038/nnano.2009.152} {\bibfield
  {journal} {\bibinfo  {journal} {Nature Nanotech}\ }\textbf {\bibinfo {volume}
  {4}},\ \bibinfo {pages} {445} (\bibinfo {year} {2009})}\BibitemShut {NoStop}%
\bibitem [{\citenamefont {Poggio}\ and\ \citenamefont
  {Degen}(2010)}]{Poggio2010}%
  \BibitemOpen
  \bibfield  {author} {\bibinfo {author} {\bibfnamefont {M.}~\bibnamefont
  {Poggio}}\ and\ \bibinfo {author} {\bibfnamefont {C.~L.}\ \bibnamefont
  {Degen}},\ }\bibfield  {title} {\bibinfo {title} {Force-detected nuclear
  magnetic resonance: Recent advances and future challenges},\ }\href
  {https://doi.org/10.1088/0957-4484/21/34/342001} {\bibfield  {journal}
  {\bibinfo  {journal} {Nanotechnology}\ }\textbf {\bibinfo {volume} {21}},\
  \bibinfo {pages} {342001} (\bibinfo {year} {2010})}\BibitemShut {NoStop}%
\bibitem [{\citenamefont {Arlett}\ \emph {et~al.}(2011)\citenamefont {Arlett},
  \citenamefont {Myers},\ and\ \citenamefont {Roukes}}]{Arlett2011}%
  \BibitemOpen
  \bibfield  {author} {\bibinfo {author} {\bibfnamefont {J.}~\bibnamefont
  {Arlett}}, \bibinfo {author} {\bibfnamefont {E.}~\bibnamefont {Myers}},\ and\
  \bibinfo {author} {\bibfnamefont {M.}~\bibnamefont {Roukes}},\ }\bibfield
  {title} {\bibinfo {title} {Comparative advantages of mechanical biosensors},\
  }\href {https://doi.org/10.1038/nnano.2011.44} {\bibfield  {journal}
  {\bibinfo  {journal} {Nature Nanotech}\ }\textbf {\bibinfo {volume} {6}},\
  \bibinfo {pages} {203} (\bibinfo {year} {2011})}\BibitemShut {NoStop}%
\bibitem [{\citenamefont {Romero}\ \emph {et~al.}(2020)\citenamefont {Romero},
  \citenamefont {Valenzuela}, \citenamefont {Kermany}, \citenamefont
  {Sementilli}, \citenamefont {Iacopi},\ and\ \citenamefont
  {Bowen}}]{Romero2020}%
  \BibitemOpen
  \bibfield  {author} {\bibinfo {author} {\bibfnamefont {E.}~\bibnamefont
  {Romero}}, \bibinfo {author} {\bibfnamefont {V.~M.}\ \bibnamefont
  {Valenzuela}}, \bibinfo {author} {\bibfnamefont {A.~R.}\ \bibnamefont
  {Kermany}}, \bibinfo {author} {\bibfnamefont {L.}~\bibnamefont {Sementilli}},
  \bibinfo {author} {\bibfnamefont {F.}~\bibnamefont {Iacopi}},\ and\ \bibinfo
  {author} {\bibfnamefont {W.~P.}\ \bibnamefont {Bowen}},\ }\bibfield  {title}
  {\bibinfo {title} {Engineering the {{Dissipation}} of {{Crystalline
  Micromechanical Resonators}}},\ }\href
  {https://doi.org/10.1103/PhysRevApplied.13.044007} {\bibfield  {journal}
  {\bibinfo  {journal} {Phys. Rev. Applied}\ }\textbf {\bibinfo {volume}
  {13}},\ \bibinfo {pages} {044007} (\bibinfo {year} {2020})}\BibitemShut
  {NoStop}%
\bibitem [{\citenamefont {Chien}\ \emph {et~al.}(2020)\citenamefont {Chien},
  \citenamefont {Steurer}, \citenamefont {Sadeghi}, \citenamefont {Cazier},\
  and\ \citenamefont {Schmid}}]{Chien2020}%
  \BibitemOpen
  \bibfield  {author} {\bibinfo {author} {\bibfnamefont {M.-H.}\ \bibnamefont
  {Chien}}, \bibinfo {author} {\bibfnamefont {J.}~\bibnamefont {Steurer}},
  \bibinfo {author} {\bibfnamefont {P.}~\bibnamefont {Sadeghi}}, \bibinfo
  {author} {\bibfnamefont {N.}~\bibnamefont {Cazier}},\ and\ \bibinfo {author}
  {\bibfnamefont {S.}~\bibnamefont {Schmid}},\ }\bibfield  {title} {\bibinfo
  {title} {Nanoelectromechanical {{Position-Sensitive Detector}} with
  {{Picometer Resolution}}},\ }\href
  {https://doi.org/10.1021/acsphotonics.0c00701} {\bibfield  {journal}
  {\bibinfo  {journal} {ACS Photonics}\ }\textbf {\bibinfo {volume} {7}},\
  \bibinfo {pages} {2197} (\bibinfo {year} {2020})}\BibitemShut {NoStop}%
\bibitem [{\citenamefont {Vicarelli}\ \emph {et~al.}(2022)\citenamefont
  {Vicarelli}, \citenamefont {Tredicucci},\ and\ \citenamefont
  {Pitanti}}]{Vicarelli2022}%
  \BibitemOpen
  \bibfield  {author} {\bibinfo {author} {\bibfnamefont {L.}~\bibnamefont
  {Vicarelli}}, \bibinfo {author} {\bibfnamefont {A.}~\bibnamefont
  {Tredicucci}},\ and\ \bibinfo {author} {\bibfnamefont {A.}~\bibnamefont
  {Pitanti}},\ }\bibfield  {title} {\bibinfo {title} {Micromechanical
  {{Bolometers}} for {{Subterahertz Detection}} at {{Room Temperature}}},\
  }\href {https://doi.org/10.1021/acsphotonics.1c01273} {\bibfield  {journal}
  {\bibinfo  {journal} {ACS Photonics}\ }\textbf {\bibinfo {volume} {9}},\
  \bibinfo {pages} {360} (\bibinfo {year} {2022})}\BibitemShut {NoStop}%
\bibitem [{\citenamefont {Manjeshwar}\ \emph {et~al.}(2023)\citenamefont
  {Manjeshwar}, \citenamefont {Ciers}, \citenamefont {Hellman}, \citenamefont
  {Bl{\"a}sing}, \citenamefont {Strittmatter},\ and\ \citenamefont
  {Wieczorek}}]{Manjeshwar2023}%
  \BibitemOpen
  \bibfield  {author} {\bibinfo {author} {\bibfnamefont {S.~K.}\ \bibnamefont
  {Manjeshwar}}, \bibinfo {author} {\bibfnamefont {A.}~\bibnamefont {Ciers}},
  \bibinfo {author} {\bibfnamefont {F.}~\bibnamefont {Hellman}}, \bibinfo
  {author} {\bibfnamefont {J.}~\bibnamefont {Bl{\"a}sing}}, \bibinfo {author}
  {\bibfnamefont {A.}~\bibnamefont {Strittmatter}},\ and\ \bibinfo {author}
  {\bibfnamefont {W.}~\bibnamefont {Wieczorek}},\ }\bibfield  {title} {\bibinfo
  {title} {High-{{Q Trampoline Resonators}} from {{Strained Crystalline InGaP}}
  for {{Integrated Free-Space Optomechanics}}},\ }\href
  {https://doi.org/10.1021/acs.nanolett.3c00996} {\bibfield  {journal}
  {\bibinfo  {journal} {Nano Lett.}\ }\textbf {\bibinfo {volume} {23}},\
  \bibinfo {pages} {5076} (\bibinfo {year} {2023})}\BibitemShut {NoStop}%
\bibitem [{\citenamefont {Reinhardt}\ \emph {et~al.}(2024)\citenamefont
  {Reinhardt}, \citenamefont {Masalehdan}, \citenamefont {Croatto},
  \citenamefont {Franke}, \citenamefont {Kunze}, \citenamefont {Schaffran},
  \citenamefont {S{\"u}ltmann}, \citenamefont {Lindner},\ and\ \citenamefont
  {Schnabel}}]{Reinhardt2024}%
  \BibitemOpen
  \bibfield  {author} {\bibinfo {author} {\bibfnamefont {C.}~\bibnamefont
  {Reinhardt}}, \bibinfo {author} {\bibfnamefont {H.}~\bibnamefont
  {Masalehdan}}, \bibinfo {author} {\bibfnamefont {S.}~\bibnamefont {Croatto}},
  \bibinfo {author} {\bibfnamefont {A.}~\bibnamefont {Franke}}, \bibinfo
  {author} {\bibfnamefont {M.~B.~K.}\ \bibnamefont {Kunze}}, \bibinfo {author}
  {\bibfnamefont {J.}~\bibnamefont {Schaffran}}, \bibinfo {author}
  {\bibfnamefont {N.}~\bibnamefont {S{\"u}ltmann}}, \bibinfo {author}
  {\bibfnamefont {A.}~\bibnamefont {Lindner}},\ and\ \bibinfo {author}
  {\bibfnamefont {R.}~\bibnamefont {Schnabel}},\ }\bibfield  {title} {\bibinfo
  {title} {Self-{{Calibrating Gas Pressure Sensor}} with a 10-{{Decade
  Measurement Range}}},\ }\href {https://doi.org/10.1021/acsphotonics.3c01488}
  {\bibfield  {journal} {\bibinfo  {journal} {ACS Photonics}\ }\textbf
  {\bibinfo {volume} {11}},\ \bibinfo {pages} {1438} (\bibinfo {year}
  {2024})}\BibitemShut {NoStop}%
\bibitem [{\citenamefont {Norte}\ \emph {et~al.}(2016)\citenamefont {Norte},
  \citenamefont {Moura},\ and\ \citenamefont {Gr{\"o}blacher}}]{Norte2016}%
  \BibitemOpen
  \bibfield  {author} {\bibinfo {author} {\bibfnamefont {R.~A.}\ \bibnamefont
  {Norte}}, \bibinfo {author} {\bibfnamefont {J.~P.}\ \bibnamefont {Moura}},\
  and\ \bibinfo {author} {\bibfnamefont {S.}~\bibnamefont {Gr{\"o}blacher}},\
  }\bibfield  {title} {\bibinfo {title} {Mechanical {{Resonators}} for
  {{Quantum Optomechanics Experiments}} at {{Room Temperature}}},\ }\href
  {https://doi.org/10.1103/PhysRevLett.116.147202} {\bibfield  {journal}
  {\bibinfo  {journal} {Phys. Rev. Lett.}\ }\textbf {\bibinfo {volume} {116}},\
  \bibinfo {pages} {147202} (\bibinfo {year} {2016})}\BibitemShut {NoStop}%
\bibitem [{\citenamefont {Ovartchaiyapong}\ \emph {et~al.}(2012)\citenamefont
  {Ovartchaiyapong}, \citenamefont {Pascal}, \citenamefont {Myers},
  \citenamefont {Lauria},\ and\ \citenamefont
  {Bleszynski~Jayich}}]{Ovartchaiyapong2012}%
  \BibitemOpen
  \bibfield  {author} {\bibinfo {author} {\bibfnamefont {P.}~\bibnamefont
  {Ovartchaiyapong}}, \bibinfo {author} {\bibfnamefont {L.~M.~A.}\ \bibnamefont
  {Pascal}}, \bibinfo {author} {\bibfnamefont {B.~A.}\ \bibnamefont {Myers}},
  \bibinfo {author} {\bibfnamefont {P.}~\bibnamefont {Lauria}},\ and\ \bibinfo
  {author} {\bibfnamefont {A.~C.}\ \bibnamefont {Bleszynski~Jayich}},\
  }\bibfield  {title} {\bibinfo {title} {High quality factor single-crystal
  diamond mechanical resonators},\ }\href {https://doi.org/10.1063/1.4760274}
  {\bibfield  {journal} {\bibinfo  {journal} {Appl. Phys. Lett.}\ }\textbf
  {\bibinfo {volume} {101}},\ \bibinfo {pages} {163505} (\bibinfo {year}
  {2012})}\BibitemShut {NoStop}%
\bibitem [{\citenamefont {Tao}\ \emph {et~al.}(2014)\citenamefont {Tao},
  \citenamefont {Boss}, \citenamefont {Moores},\ and\ \citenamefont
  {Degen}}]{Tao2014}%
  \BibitemOpen
  \bibfield  {author} {\bibinfo {author} {\bibfnamefont {Y.}~\bibnamefont
  {Tao}}, \bibinfo {author} {\bibfnamefont {J.~M.}\ \bibnamefont {Boss}},
  \bibinfo {author} {\bibfnamefont {B.~A.}\ \bibnamefont {Moores}},\ and\
  \bibinfo {author} {\bibfnamefont {C.~L.}\ \bibnamefont {Degen}},\ }\bibfield
  {title} {\bibinfo {title} {Single-crystal diamond nanomechanical resonators
  with quality factors exceeding one million},\ }\href
  {https://doi.org/10.1038/ncomms4638} {\bibfield  {journal} {\bibinfo
  {journal} {Nat Commun}\ }\textbf {\bibinfo {volume} {5}},\ \bibinfo {pages}
  {3638} (\bibinfo {year} {2014})}\BibitemShut {NoStop}%
\bibitem [{\citenamefont {H{\'e}ritier}\ \emph {et~al.}(2018)\citenamefont
  {H{\'e}ritier}, \citenamefont {Eichler}, \citenamefont {Pan}, \citenamefont
  {Grob}, \citenamefont {Shorubalko}, \citenamefont {Krass}, \citenamefont
  {Tao},\ and\ \citenamefont {Degen}}]{Heritier2018}%
  \BibitemOpen
  \bibfield  {author} {\bibinfo {author} {\bibfnamefont {M.}~\bibnamefont
  {H{\'e}ritier}}, \bibinfo {author} {\bibfnamefont {A.}~\bibnamefont
  {Eichler}}, \bibinfo {author} {\bibfnamefont {Y.}~\bibnamefont {Pan}},
  \bibinfo {author} {\bibfnamefont {U.}~\bibnamefont {Grob}}, \bibinfo {author}
  {\bibfnamefont {I.}~\bibnamefont {Shorubalko}}, \bibinfo {author}
  {\bibfnamefont {M.~D.}\ \bibnamefont {Krass}}, \bibinfo {author}
  {\bibfnamefont {Y.}~\bibnamefont {Tao}},\ and\ \bibinfo {author}
  {\bibfnamefont {C.~L.}\ \bibnamefont {Degen}},\ }\bibfield  {title} {\bibinfo
  {title} {Nanoladder {{Cantilevers Made}} from {{Diamond}} and {{Silicon}}},\
  }\href {https://doi.org/10.1021/acs.nanolett.7b05035} {\bibfield  {journal}
  {\bibinfo  {journal} {Nano Lett.}\ }\textbf {\bibinfo {volume} {18}},\
  \bibinfo {pages} {1814} (\bibinfo {year} {2018})}\BibitemShut {NoStop}%
\bibitem [{\citenamefont {Xu}\ \emph {et~al.}(2024)\citenamefont {Xu},
  \citenamefont {Shin}, \citenamefont {Sberna}, \citenamefont {{van der Kolk}},
  \citenamefont {Cupertino}, \citenamefont {Bessa},\ and\ \citenamefont
  {Norte}}]{Xu2024}%
  \BibitemOpen
  \bibfield  {author} {\bibinfo {author} {\bibfnamefont {M.}~\bibnamefont
  {Xu}}, \bibinfo {author} {\bibfnamefont {D.}~\bibnamefont {Shin}}, \bibinfo
  {author} {\bibfnamefont {P.~M.}\ \bibnamefont {Sberna}}, \bibinfo {author}
  {\bibfnamefont {R.}~\bibnamefont {{van der Kolk}}}, \bibinfo {author}
  {\bibfnamefont {A.}~\bibnamefont {Cupertino}}, \bibinfo {author}
  {\bibfnamefont {M.~A.}\ \bibnamefont {Bessa}},\ and\ \bibinfo {author}
  {\bibfnamefont {R.~A.}\ \bibnamefont {Norte}},\ }\bibfield  {title} {\bibinfo
  {title} {High-{{Strength Amorphous Silicon Carbide}} for {{Nanomechanics}}},\
  }\href {https://doi.org/10.1002/adma.202306513} {\bibfield  {journal}
  {\bibinfo  {journal} {Adv. Mater.}\ }\textbf {\bibinfo {volume} {36}},\
  \bibinfo {pages} {2306513} (\bibinfo {year} {2024})}\BibitemShut {NoStop}%
\bibitem [{\citenamefont {Zubko}\ \emph {et~al.}(2011)\citenamefont {Zubko},
  \citenamefont {Gariglio}, \citenamefont {Gabay}, \citenamefont {Ghosez},\
  and\ \citenamefont {Triscone}}]{Zubko2011}%
  \BibitemOpen
  \bibfield  {author} {\bibinfo {author} {\bibfnamefont {P.}~\bibnamefont
  {Zubko}}, \bibinfo {author} {\bibfnamefont {S.}~\bibnamefont {Gariglio}},
  \bibinfo {author} {\bibfnamefont {M.}~\bibnamefont {Gabay}}, \bibinfo
  {author} {\bibfnamefont {P.}~\bibnamefont {Ghosez}},\ and\ \bibinfo {author}
  {\bibfnamefont {J.-M.}\ \bibnamefont {Triscone}},\ }\bibfield  {title}
  {\bibinfo {title} {Interface {{Physics}} in {{Complex Oxide
  Heterostructures}}},\ }\href
  {https://doi.org/10.1146/annurev-conmatphys-062910-140445} {\bibfield
  {journal} {\bibinfo  {journal} {Annu. Rev. Condens. Matter Phys.}\ }\textbf
  {\bibinfo {volume} {2}},\ \bibinfo {pages} {141} (\bibinfo {year}
  {2011})}\BibitemShut {NoStop}%
\bibitem [{\citenamefont {Manca}\ \emph
  {et~al.}(2024{\natexlab{a}})\citenamefont {Manca}, \citenamefont
  {Kalaboukhov}, \citenamefont {Plaza}, \citenamefont {Cichetto}, \citenamefont
  {Wahlberg}, \citenamefont {Bellingeri}, \citenamefont {Bisio}, \citenamefont
  {Lombardi}, \citenamefont {Marr{\'e}},\ and\ \citenamefont
  {Pellegrino}}]{Manca2024a}%
  \BibitemOpen
  \bibfield  {author} {\bibinfo {author} {\bibfnamefont {N.}~\bibnamefont
  {Manca}}, \bibinfo {author} {\bibfnamefont {A.}~\bibnamefont {Kalaboukhov}},
  \bibinfo {author} {\bibfnamefont {A.~E.}\ \bibnamefont {Plaza}}, \bibinfo
  {author} {\bibfnamefont {L.}~\bibnamefont {Cichetto}}, \bibinfo {author}
  {\bibfnamefont {E.}~\bibnamefont {Wahlberg}}, \bibinfo {author}
  {\bibfnamefont {E.}~\bibnamefont {Bellingeri}}, \bibinfo {author}
  {\bibfnamefont {F.}~\bibnamefont {Bisio}}, \bibinfo {author} {\bibfnamefont
  {F.}~\bibnamefont {Lombardi}}, \bibinfo {author} {\bibfnamefont
  {D.}~\bibnamefont {Marr{\'e}}},\ and\ \bibinfo {author} {\bibfnamefont
  {L.}~\bibnamefont {Pellegrino}},\ }\bibfield  {title} {\bibinfo {title}
  {Integration of {{High}}-{{Tc Superconductors}} with {{High}}-{{Q}}-{{Factor
  Oxide Mechanical Resonators}}},\ }\href
  {https://doi.org/10.1002/adfm.202403155} {\bibfield  {journal} {\bibinfo
  {journal} {Adv Funct Materials}\ ,\ \bibinfo {pages} {2403155}} (\bibinfo
  {year} {2024}{\natexlab{a}})}\BibitemShut {NoStop}%
\bibitem [{\citenamefont {Forstner}\ \emph {et~al.}(2012)\citenamefont
  {Forstner}, \citenamefont {Prams}, \citenamefont {Knittel}, \citenamefont
  {{van Ooijen}}, \citenamefont {Swaim}, \citenamefont {Harris}, \citenamefont
  {Szorkovszky}, \citenamefont {Bowen},\ and\ \citenamefont
  {{Rubinsztein-Dunlop}}}]{Forstner2012}%
  \BibitemOpen
  \bibfield  {author} {\bibinfo {author} {\bibfnamefont {S.}~\bibnamefont
  {Forstner}}, \bibinfo {author} {\bibfnamefont {S.}~\bibnamefont {Prams}},
  \bibinfo {author} {\bibfnamefont {J.}~\bibnamefont {Knittel}}, \bibinfo
  {author} {\bibfnamefont {E.~D.}\ \bibnamefont {{van Ooijen}}}, \bibinfo
  {author} {\bibfnamefont {J.~D.}\ \bibnamefont {Swaim}}, \bibinfo {author}
  {\bibfnamefont {G.~I.}\ \bibnamefont {Harris}}, \bibinfo {author}
  {\bibfnamefont {A.}~\bibnamefont {Szorkovszky}}, \bibinfo {author}
  {\bibfnamefont {W.~P.}\ \bibnamefont {Bowen}},\ and\ \bibinfo {author}
  {\bibfnamefont {H.}~\bibnamefont {{Rubinsztein-Dunlop}}},\ }\bibfield
  {title} {\bibinfo {title} {Cavity {{Optomechanical Magnetometer}}},\ }\href
  {https://doi.org/10.1103/PhysRevLett.108.120801} {\bibfield  {journal}
  {\bibinfo  {journal} {Phys. Rev. Lett.}\ }\textbf {\bibinfo {volume} {108}},\
  \bibinfo {pages} {120801} (\bibinfo {year} {2012})}\BibitemShut {NoStop}%
\bibitem [{\citenamefont {Fischer}\ \emph {et~al.}(2019)\citenamefont
  {Fischer}, \citenamefont {McNally}, \citenamefont {Reetz}, \citenamefont
  {Assump{\c c}{\~a}o}, \citenamefont {Knief}, \citenamefont {Lin},\ and\
  \citenamefont {Regal}}]{Fischer2019}%
  \BibitemOpen
  \bibfield  {author} {\bibinfo {author} {\bibfnamefont {R.}~\bibnamefont
  {Fischer}}, \bibinfo {author} {\bibfnamefont {D.~P.}\ \bibnamefont
  {McNally}}, \bibinfo {author} {\bibfnamefont {C.}~\bibnamefont {Reetz}},
  \bibinfo {author} {\bibfnamefont {G.~G.~T.}\ \bibnamefont {Assump{\c
  c}{\~a}o}}, \bibinfo {author} {\bibfnamefont {T.}~\bibnamefont {Knief}},
  \bibinfo {author} {\bibfnamefont {Y.}~\bibnamefont {Lin}},\ and\ \bibinfo
  {author} {\bibfnamefont {C.~A.}\ \bibnamefont {Regal}},\ }\bibfield  {title}
  {\bibinfo {title} {Spin detection with a micromechanical trampoline: Towards
  magnetic resonance microscopy harnessing cavity optomechanics},\ }\href
  {https://doi.org/10.1088/1367-2630/ab117a} {\bibfield  {journal} {\bibinfo
  {journal} {New J. Phys.}\ }\textbf {\bibinfo {volume} {21}},\ \bibinfo
  {pages} {043049} (\bibinfo {year} {2019})}\BibitemShut {NoStop}%
\bibitem [{\citenamefont {Maspero}\ \emph {et~al.}(2021)\citenamefont
  {Maspero}, \citenamefont {Gatani}, \citenamefont {Cuccurullo},\ and\
  \citenamefont {Bertacco}}]{Maspero2021}%
  \BibitemOpen
  \bibfield  {author} {\bibinfo {author} {\bibfnamefont {F.}~\bibnamefont
  {Maspero}}, \bibinfo {author} {\bibfnamefont {G.}~\bibnamefont {Gatani}},
  \bibinfo {author} {\bibfnamefont {S.}~\bibnamefont {Cuccurullo}},\ and\
  \bibinfo {author} {\bibfnamefont {R.}~\bibnamefont {Bertacco}},\ }\bibfield
  {title} {\bibinfo {title} {{{MEMS Magnetometer Using Magnetic Flux
  Concentrators}} and {{Permanent Magnets}}},\ }in\ \href
  {https://doi.org/10.1109/MEMS51782.2021.9375441} {\emph {\bibinfo {booktitle}
  {2021 {{IEEE}} 34th {{Int}}. {{Conf}}. {{Micro Electro Mech}}. {{Syst}}.
  {{MEMS}}}}}\ (\bibinfo  {publisher} {IEEE},\ \bibinfo {address} {Gainesville,
  FL, USA},\ \bibinfo {year} {2021})\ pp.\ \bibinfo {pages}
  {374--377}\BibitemShut {NoStop}%
\bibitem [{\citenamefont {Pellegrino}\ \emph {et~al.}(2021)\citenamefont
  {Pellegrino}, \citenamefont {Manca}, \citenamefont {Marre'}, \citenamefont
  {Remaggi}, \citenamefont {Bertacco}, \citenamefont {Maspero}, \citenamefont
  {Venstra}, \citenamefont {{della Penna}}, \citenamefont {Hilschenz},
  \citenamefont {Kalaboukhov},\ and\ \citenamefont
  {Lombardi}}]{Pellegrino2021a}%
  \BibitemOpen
  \bibfield  {author} {\bibinfo {author} {\bibfnamefont {L.}~\bibnamefont
  {Pellegrino}}, \bibinfo {author} {\bibfnamefont {N.}~\bibnamefont {Manca}},
  \bibinfo {author} {\bibfnamefont {D.}~\bibnamefont {Marre'}}, \bibinfo
  {author} {\bibfnamefont {F.}~\bibnamefont {Remaggi}}, \bibinfo {author}
  {\bibfnamefont {R.}~\bibnamefont {Bertacco}}, \bibinfo {author}
  {\bibfnamefont {F.}~\bibnamefont {Maspero}}, \bibinfo {author} {\bibfnamefont
  {W.~J.}\ \bibnamefont {Venstra}}, \bibinfo {author} {\bibfnamefont
  {S.}~\bibnamefont {{della Penna}}}, \bibinfo {author} {\bibfnamefont
  {I.}~\bibnamefont {Hilschenz}}, \bibinfo {author} {\bibfnamefont
  {A.}~\bibnamefont {Kalaboukhov}},\ and\ \bibinfo {author} {\bibfnamefont
  {F.}~\bibnamefont {Lombardi}},\ }\href
  {https://patents.google.com/patent/EP3896470A1/en?oq=EP3896470A1} {\bibinfo
  {title} {A device for sensing a magnetic field}} (\bibinfo {year}
  {2021})\BibitemShut {NoStop}%
\bibitem [{\citenamefont {Berndt}\ \emph {et~al.}(2000)\citenamefont {Berndt},
  \citenamefont {Balbarin},\ and\ \citenamefont {Suzuki}}]{Berndt2000}%
  \BibitemOpen
  \bibfield  {author} {\bibinfo {author} {\bibfnamefont {L.~M.}\ \bibnamefont
  {Berndt}}, \bibinfo {author} {\bibfnamefont {V.}~\bibnamefont {Balbarin}},\
  and\ \bibinfo {author} {\bibfnamefont {Y.}~\bibnamefont {Suzuki}},\
  }\bibfield  {title} {\bibinfo {title} {Magnetic anisotropy and strain states
  of (001) and (110) colossal magnetoresistance thin films},\ }\href
  {https://doi.org/10.1063/1.1321733} {\bibfield  {journal} {\bibinfo
  {journal} {Appl. Phys. Lett.}\ }\textbf {\bibinfo {volume} {77}},\ \bibinfo
  {pages} {2903} (\bibinfo {year} {2000})}\BibitemShut {NoStop}%
\bibitem [{\citenamefont {Manca}\ \emph {et~al.}(2022)\citenamefont {Manca},
  \citenamefont {Remaggi}, \citenamefont {Plaza}, \citenamefont {Varbaro},
  \citenamefont {Bernini}, \citenamefont {Pellegrino},\ and\ \citenamefont
  {Marr{\'e}}}]{Manca2022}%
  \BibitemOpen
  \bibfield  {author} {\bibinfo {author} {\bibfnamefont {N.}~\bibnamefont
  {Manca}}, \bibinfo {author} {\bibfnamefont {F.}~\bibnamefont {Remaggi}},
  \bibinfo {author} {\bibfnamefont {A.~E.}\ \bibnamefont {Plaza}}, \bibinfo
  {author} {\bibfnamefont {L.}~\bibnamefont {Varbaro}}, \bibinfo {author}
  {\bibfnamefont {C.}~\bibnamefont {Bernini}}, \bibinfo {author} {\bibfnamefont
  {L.}~\bibnamefont {Pellegrino}},\ and\ \bibinfo {author} {\bibfnamefont
  {D.}~\bibnamefont {Marr{\'e}}},\ }\bibfield  {title} {\bibinfo {title}
  {Stress {{Analysis}} and {{Q}}-{{Factor}} of {{Free}}-{{Standing}}
  ({{La}},{{Sr}}){{MnO}}{\textsubscript{3}} {{Oxide Resonators}}},\ }\href
  {https://doi.org/10.1002/smll.202202768} {\bibfield  {journal} {\bibinfo
  {journal} {Small}\ }\textbf {\bibinfo {volume} {18}},\ \bibinfo {pages}
  {2202768} (\bibinfo {year} {2022})}\BibitemShut {NoStop}%
\bibitem [{\citenamefont {Ceriale}\ \emph {et~al.}(2014)\citenamefont
  {Ceriale}, \citenamefont {Pellegrino}, \citenamefont {Manca},\ and\
  \citenamefont {Marr{\'e}}}]{Ceriale2014}%
  \BibitemOpen
  \bibfield  {author} {\bibinfo {author} {\bibfnamefont {V.}~\bibnamefont
  {Ceriale}}, \bibinfo {author} {\bibfnamefont {L.}~\bibnamefont {Pellegrino}},
  \bibinfo {author} {\bibfnamefont {N.}~\bibnamefont {Manca}},\ and\ \bibinfo
  {author} {\bibfnamefont {D.}~\bibnamefont {Marr{\'e}}},\ }\bibfield  {title}
  {\bibinfo {title} {Electro-thermal bistability in
  ({{La}}{\textsubscript{0.7}}{{Sr}}{\textsubscript{0.3}}){{MnO}}{\textsubscript{3}}
  suspended microbridges: {{Thermal}} characterization and transient
  analysis},\ }\href {https://doi.org/10.1063/1.4864222} {\bibfield  {journal}
  {\bibinfo  {journal} {Journal of Applied Physics}\ }\textbf {\bibinfo
  {volume} {115}},\ \bibinfo {pages} {054511} (\bibinfo {year}
  {2014})}\BibitemShut {NoStop}%
\bibitem [{\citenamefont {Plaza}\ \emph {et~al.}(2021)\citenamefont {Plaza},
  \citenamefont {Manca}, \citenamefont {Bernini}, \citenamefont {Marr{\'e}},\
  and\ \citenamefont {Pellegrino}}]{Plaza2021}%
  \BibitemOpen
  \bibfield  {author} {\bibinfo {author} {\bibfnamefont {A.~E.}\ \bibnamefont
  {Plaza}}, \bibinfo {author} {\bibfnamefont {N.}~\bibnamefont {Manca}},
  \bibinfo {author} {\bibfnamefont {C.}~\bibnamefont {Bernini}}, \bibinfo
  {author} {\bibfnamefont {D.}~\bibnamefont {Marr{\'e}}},\ and\ \bibinfo
  {author} {\bibfnamefont {L.}~\bibnamefont {Pellegrino}},\ }\bibfield  {title}
  {\bibinfo {title} {The role of etching anisotropy in the fabrication of
  freestanding oxide microstructures on {{SrTiO}}{\textsubscript{3}}(100),
  {{SrTiO}}{\textsubscript{3}}(110), and {{SrTiO}}{\textsubscript{3}}(111)
  substrates},\ }\href {https://doi.org/10.1063/5.0056524} {\bibfield
  {journal} {\bibinfo  {journal} {Appl. Phys. Lett.}\ }\textbf {\bibinfo
  {volume} {119}},\ \bibinfo {pages} {033504} (\bibinfo {year}
  {2021})}\BibitemShut {NoStop}%
\bibitem [{\citenamefont {Manca}\ \emph
  {et~al.}(2024{\natexlab{b}})\citenamefont {Manca}, \citenamefont {Plaza},
  \citenamefont {Jr}, \citenamefont {Venstra}, \citenamefont {Bernini},
  \citenamefont {Marr{\'e}},\ and\ \citenamefont {Pellegrino}}]{Manca2024}%
  \BibitemOpen
  \bibfield  {author} {\bibinfo {author} {\bibfnamefont {N.}~\bibnamefont
  {Manca}}, \bibinfo {author} {\bibfnamefont {A.~E.}\ \bibnamefont {Plaza}},
  \bibinfo {author} {\bibfnamefont {L.~C.}\ \bibnamefont {Jr}}, \bibinfo
  {author} {\bibfnamefont {W.~J.}\ \bibnamefont {Venstra}}, \bibinfo {author}
  {\bibfnamefont {C.}~\bibnamefont {Bernini}}, \bibinfo {author} {\bibfnamefont
  {D.}~\bibnamefont {Marr{\'e}}},\ and\ \bibinfo {author} {\bibfnamefont
  {L.}~\bibnamefont {Pellegrino}},\ }\href {http://arxiv.org/abs/2410.06758}
  {\bibinfo {title} {Oxide {{Membranes}} from {{Bulk Micro-Machining}} of
  {{SrTiO3}} substrates}} (\bibinfo {year} {2024}{\natexlab{b}}),\ \Eprint
  {https://arxiv.org/abs/2410.06758} {arXiv:2410.06758 [cond-mat]} \BibitemShut
  {NoStop}%
\bibitem [{\citenamefont {Tokura}\ and\ \citenamefont
  {Tomioka}(1999)}]{Tokura1999}%
  \BibitemOpen
  \bibfield  {author} {\bibinfo {author} {\bibfnamefont {Y.}~\bibnamefont
  {Tokura}}\ and\ \bibinfo {author} {\bibfnamefont {Y.}~\bibnamefont
  {Tomioka}},\ }\bibfield  {title} {\bibinfo {title} {Colossal magnetoresistive
  manganites},\ }\href {https://doi.org/10.1016/S0304-8853(99)00352-2}
  {\bibfield  {journal} {\bibinfo  {journal} {J. Magn. Magn. Mater.}\ ,\
  \bibinfo {pages} {23}} (\bibinfo {year} {1999})}\BibitemShut {NoStop}%
\bibitem [{\citenamefont {Mitchell}\ \emph {et~al.}(1996)\citenamefont
  {Mitchell}, \citenamefont {Argyriou}, \citenamefont {Potter}, \citenamefont
  {Hinks}, \citenamefont {Jorgensen},\ and\ \citenamefont
  {Bader}}]{Mitchell1996}%
  \BibitemOpen
  \bibfield  {author} {\bibinfo {author} {\bibfnamefont {J.~F.}\ \bibnamefont
  {Mitchell}}, \bibinfo {author} {\bibfnamefont {D.~N.}\ \bibnamefont
  {Argyriou}}, \bibinfo {author} {\bibfnamefont {C.~D.}\ \bibnamefont
  {Potter}}, \bibinfo {author} {\bibfnamefont {D.~G.}\ \bibnamefont {Hinks}},
  \bibinfo {author} {\bibfnamefont {J.~D.}\ \bibnamefont {Jorgensen}},\ and\
  \bibinfo {author} {\bibfnamefont {S.~D.}\ \bibnamefont {Bader}},\ }\bibfield
  {title} {\bibinfo {title} {Structural phase diagram of
  {{La}}{\textsubscript{(1-x)}}{{Sr}}{\textsubscript{x}}{{MnO}}{\textsubscript{3+{$\delta$}}}:
  {{Relationship}} to magnetic and transport properties},\ }\href
  {https://doi.org/10.1103/PhysRevB.54.6172} {\bibfield  {journal} {\bibinfo
  {journal} {Phys. Rev. B}\ }\textbf {\bibinfo {volume} {54}},\ \bibinfo
  {pages} {6172} (\bibinfo {year} {1996})}\BibitemShut {NoStop}%
\end{thebibliography}%


\newpage\newpage

\foreach \x in {1,...,3}
{
	\clearpage
	\includepdf[pages={\x}]{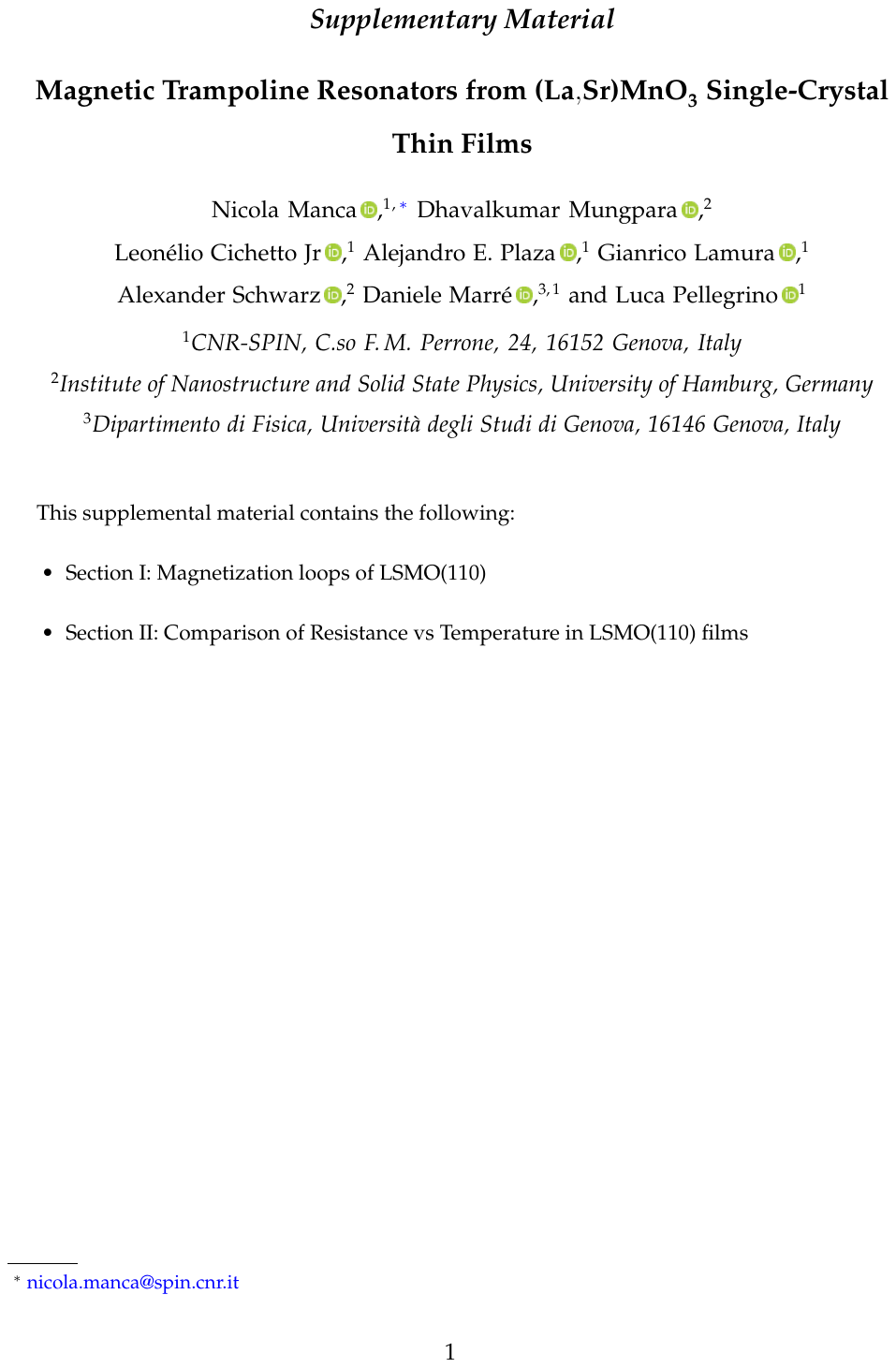}
}

\end{document}